# "Results concerning the centre of our galaxy"

J. Dunning-Davies,
University of Hull,
Hull HU67RX,
England.

J.Dunning-Davies@hull.ac.uk

For many years it was felt that, when a star collapsed, a white dwarf resulted if the mass of the original star was below the Chandrasekhar limit, a neutron star if the mass was somewhat larger but still less than four or five solar masses, but after that black holes were felt to provide the only possible final state. The extension of this hierarchy to include the possibility of quark, and even sub-quark, stars has been proposed and here is used to offer an alternative explanation for the recently published photograph, credited to Eckart and Genzel, purporting to show stars near the centre of our Galaxy moving at very high speeds. The same basic ideas are used also to consider the even more recent results of Schödel and collaborators concerning the detailed observations of a stellar orbit very close to the centre of our Galaxy.

. **1. Introduction.**

Recently, a photograph, credited to Eckart and Genzel [1], apparently showing stars near the centre of our galaxy moving at very high speeds, was released. In a possible explanation offered for this occurrence, the presence of a massive, compact central object was claimed. It was also claimed that further analysis indicated a mass equivalent to more than a million solar masses was confined to region of radius less than one tenth of a light year. Astronomers are said to have interpreted all this as strong evidence for a massive black hole occupying the centre of our galaxy.

Even more recently, Schödel, et al [2] have reported some extremely interesting observations that could help improve knowledge of the centre of our Galaxy, and possibly even of other galaxies also. The observations in question concerned a stellar orbit around Sagittarius A*, a mere $1.8 \times 10^{13}$m from the centre of the Galaxy. The orbit implies the presence of a central gravitating mass which is found, by applying Kepler's third law, to be $3.7 \times 10^{6}$ solar masses, or $7.36 \times 10^{36}$kg. This scenario has been interpreted also as establishing the definite presence of a black hole at the Galactic centre.

Here the idea [3] that, when stars collapse in on themselves, the end result is to produce a hierarchy of objects starting with white dwarfs, which are essentially composed of degenerate electrons, via neutron stars which are composed of degenerate neutrons, to quark stars composed of degenerate quarks, is reintroduced. Such a hierarchy might well be extended to include the possibility of sub-quark stars also and eventually lead to black holes as a limiting case which may, or may not, be achieved in practice. This idea gains credence with the announcement that observations made by the orbiting Chandra X-ray Observatory [4] have indicated a possible first sighting of quark stars. These ideas are used here to offer an alternative explanation for the observations of both Eckart and Genzel and Schödel et al and also to show where doubt may be cast on the previous explanations offered. Consideration of these ideas is made even more topical by the fairly regular appearance these days of articles in popular science journals which claim that black holes have been identified; a typical example being 'The Milky Way's Dark Starving Pit' which appeared in Science of May 30[th], 2003. Further evidence for the need of such considerations is provided by the frequent articles claiming certain celestial bodies to definitely be black holes, when such absolute identification is not available.[(see5)]

In section 2, some well known results applying to black holes are reviewed and the less well known limit for the mass/radius ratio of a black hole is derived. This ratio is then applied to the situation envisaged by Eckart and Genzel and by Schödel et al and it is shown where doubt could arise over the explanations of their observations. Section 3 is then devoted to a brief résumé of results pertaining to quark stars, which are used in section 4 to provide an alternative explanation for the data under discussion.

**2. The mass/radius relation for a black hole and its consequences.**

For the case of a spherically symmetric field produced by a spherically symmetric body at rest, the Einstein equations yield the well-known Schwarzschild solution[(6)]:

$$ds^2 = \left(1 - \frac{2GM}{Rc^2}\right)dt^2 - \left(1 - \frac{2GM}{Rc^2}\right)^{-1} dR^2 - R^2 d\vartheta^2 - R^2 \sin^2\vartheta d\phi^2,$$

where $G$ is the universal constant of gravitation and $c$ is the speed of light. For this solution to be real,

$$1 - \frac{2GM}{Rc^2} > 0,$$

that is

$$M/R < 6.7 \times 10^{26} \text{kg/m}.$$

For a black hole, this inequality is not satisfied and

$$M/R \geq 6.7 \times 10^{26} \text{kg/m}.$$

It is interesting to note that this is precisely the expression for the ratio of mass to radius that Michell derived, in 1784[(7)], using purely Newtonian methods, for a body possessing an escape speed greater than, or equal to, that of light.

In a recent article, Eckart and Genzel[(1)] considered stars near the centre of our own galaxy moving at very high speeds. The explanation proffered claimed that 'if these fast stars are held to the Galactic Centre by gravity, then the central object exerting this gravity must be both compact and massive.' Further, it was claimed that 'analysis of the stellar motions indicates that, over one million times the mass of our sun is somehow confined to a region less than one fifth of a light-year across.' Astronomers are said to have interpreted 'these observations as strong evidence that the centre of our Galaxy is home to a very massive black hole.' However, is this feasible?

For a star of the size imagined, its mass would be

$$M = 10^6 \times 2 \times 10^{30} = 2 \times 10^{36} \text{kg}.$$

Hence, using the above mass/radius relation for a black hole, the star envisaged would need to have a radius

$$R = (7 \times 10^{26})^{-1} \times 2 \times 10^{36} \approx 3 \times 10^{9} \text{m}.$$

This radius is readily seen to be appreciably less than the proposal that the mass has to be confined to a region less than one fifth of a light-year across; that is, it has to be confined to a region with radius less than $10^{15}$ m. Before such a body could be claimed to be a black hole with any real degree of confidence, its mass would have to be shown confined to a region whose radius is appreciably less than $10^{15}$ m. Alternatively, the above mass/radius relation shows that, if the mass is confined to a region of radius of the order of one fifth of a light year, that mass would have to be of the order of a million million solar masses.

As far as the more recent article by Schödel et al [2] is concerned, the massive object is found to have a mass of approximately $7.36 \times 10^{36}$ kg. For such an object to be a black hole, the above relation indicates that its radius would need to be approximately $1.1 \times 10^{10}$ m, or about 37 light seconds. The measured radius of 17 light hours, although small, is three orders of magnitude greater than the predicted radius of the event horizon for the observationally inferred mass. However, in this case, the appeal to Kepler's third law involves the assumption that the central mass is essentially a point. Obviously, more information is required about the mass distribution itself. The observational results obtained this far are indeed impressive; so much so that it might appear churlish to require details for orbits of even smaller radius, but that is the situation.

### 3. Quark stars?

In the late 90's [3], it was predicted that, when stars collapse in on themselves, the end result is a hierarchy of objects starting with white dwarfs and continuing via neutron stars to quark stars and even, possibly, to sub-quark stars. Such a hierarchy was also predicted to have black holes as a limiting case, - a limit which might, or might not, be achieved. However, if achieved, such a limiting case would be a definite physical entity rather than a purely mathematical singularity. Using arguments based on energy considerations, elementary calculations of the radii of such sub-neutron bodies were made. This procedure was adopted in the absence of a full theoretical description of quark structures. It was hypothesised that a neutron would become destabilised when the external gravitational field exceeds the self-energy of the neutron. For this to be possible, the minimum energy per particle would have to be

$$\varepsilon(n) \approx 1.5 \times 10^{-10} \text{J}$$

and, for a body of four solar masses, this would be associated with a radius of about 10 km. However, the quarks might be expected to be set free by a greater energy field and the mature body might be expected to involve an energy of about $10^{-9}$ J. For a body of five solar masses, this would result in a radius of less than 10 km.

However, it has been suggested [8] that quarks themselves might be composed of particles of mass $10^{-39}$ kg and, if this were so, the body composed of such particles would possess a radius of the order of $10^{-2}$ m.

A final point worth noting is that, in all cases of quark or sub-quark stars, the escape speed is less than that of light and it is this which provoked the suggestion that a black hole might represent the limiting case of the proposed hierarchy.

## 4. An alternative explanation for the results of Eckart and Genzel.

Instead of the scenario suggested by Eckart and Genzel, consider the situation where, instead of the observed effects being assumed the result of the presence of a black hole of mass equal to a million solar masses, they are taken as being due to the presence of a million stars of solar mass, each with radius less than 3 kms. If the black hole condition referred to in section 2 holds, the radius of the volume under consideration would need to be $3 \times 10^9$ m. and, if a total of $10^6$ stars is involved, it would follow that the distance between neighbouring stars would be approximately $3 \times 10^7$ m. This would lead to a potential energy of $(6.67 \times 10^{-11} \times 4 \times 10^{60})/(3 \times 10^7)$ and, if this is put equal to the kinetic energy, it leads to a velocity given by

$$v^2 = (6.67 \times 10^{-11} \times 4 \times 10^{60})/(3 \times 10^7 \times 10^{30}) \Rightarrow v \approx 3 \times 10^6 \text{ m}.$$

This approximate value of the velocity of a star in the model under discussion is seen to be of the same order of magnitude as those noted by Eckart and Genzel. Hence, the model proposed here to describe the situation viewed by Eckart and Genzel, but with an overall radius of the order of $10^9$ m instead of a light year, would seem to lead to acceptable results. If, however, the volume under consideration is of radius of the order of $10^{15}$ m as suggested by Eckart and Genzel, then, not only is the black hole condition not satisfied, this alternative model would lead to possible star velocities of the order of $3 \times 10^3$ m.

## 5. Conclusion.

Since the arguments contained in section 2 above cast doubts over the explanation offered for the observations of Eckart and Genzel, what alternatives can be advanced? Firstly, following the ideas contained in section 3, one possible alternative could be the presence of a significant number of quark and/or sub-quark stars clustered near the centre of our Galaxy. Such an explanation gains some credence from the order of magnitude calculations of section 4 as well as from the recent announcement of the possible sighting of a quark star.

Referring to the paper by Schödel et al, a similar explanation for the results cannot be ruled out totally. Alternatively, the central mass could well be composed of a mixture of baryonic and dark matter. Such a mixture could involve a relatively small number of normal stars, of essentially solar mass, contained within a distributed source of gravitation able to constrain the mixture within a stable limited volume forming the Galactic centre. If such an identification proved true, it could provide a method for estimating the ratio of ordinary to dark matter in one particular case.

Again, considering the pictures taken by the orbiting Infrared Space Observatory telescope in 1996 [9], the idea that the centre of our Galaxy is occupied predominantly by a very large number of hot, bright young stars must also be considered.